Suggested Title: **Simultaneous optogenetic manipulation and calcium imaging in freely moving *C. elegans***

Frederick B. Shipley[1], Christopher M. Clark[2], Mark J. Alkema[2], Andrew M. Leifer[1*]
[1] Lewis Sigler Institute for Integrative Genomics, Princeton University, Princeton, NJ, USA
[2] Department of Neurobiology, University of Massachusetts Medical School, Worcester, MA, USA
* To whom correspondence should be addressed: leifer@princeton.edu

Editor:

A fundamental goal of systems neuroscience is to probe the dynamics of neural activity that generate behavior. Here we present an instrument to simultaneously manipulate and monitor neural activity and behavior in the freely moving nematode *Caenorhabditis elegans*. We use the instrument to directly observe the relationship between sensory neuron activation, interneuron dynamics and locomotion in the mechanosensory circuit.

Previously in this journal, we presented an optogenetic illumination system capable of real-time light delivery with high spatial resolution to stimulate or inhibit specified targets in freely moving *C. elegans* [1]. This "Colbert" system and others like it [2] have been instrumental in defining neural coding of several behaviors in *C. elegans* including chemotaxis [3], nociception [4] and the escape response [5]. Here we integrate the Colbert system with simultaneous monitoring of intracellular calcium activity. To our knowledge this is the first instrument of its kind. The integration of optogenetics, calcium imaging and behavioral analysis allows us to dissect neural circuit dynamics and correlate neural activity with behavior.

In this system, custom computer vision software tracks animal behavior and identifies the location of targeted neurons in real-time. The software adjusts mirrors on a digital micromirror device (DMD) to project patterned illumination onto targeted neurons. We induce neural activity using Channelrhodopsin (ChR2) and monitor calcium dynamics by simultaneously measuring the fluorescence of an optical calcium indicator, GCaMP3, and a calcium-insensitive reference, mCherry. By patterning our illumination to independently target ChR2 and GCaMP3 expressing neurons, the system can continuously observe calcium dynamics from one neuron while transiently activating others (see Fig 1a and Supplementary methods). Unlike previous investigations in immobilized animals [6], here the manipulation and analysis of neural activity was directly correlated to behavioral output.

We used the instrument to observe calcium transients in the backward locomotion command interneurons AVA, in response to activation of the anterior mechanosensory neurons ALM, AVM or both (Fig. 1b). Transgenic worms that co-expressed ChR2 in the mechanosensory neurons and a GCaMP3::SL2::mCherry operon in interneurons AVA were stimulated for 2.7s to induce a reversal (see representative trial, Fig 1c, Supplementary Fig 1 and Supplementary Movie 1). We found that AVA's calcium transients correlated with the worm's mean velocity (Pearson's correlation coefficient $R$=-0.83), irrespective of stimuli (Fig 1e). Averaging across trials, neither AVA's mean activity, the worm's mean reversal velocity nor the probability of a reversal varied significantly with stimulus (Fig 1d,e,g).

Our system illuminates cell bodies independently provided they are sufficiently spatially separated. We demonstrate independent illumination of neurons as close as 50 microns apart in a moving worm (see Supplementary Movie 2), and estimate our actual resolution to be ~30 microns. The system's effective resolution would be further improved using next generation red-shifted calcium indicators.

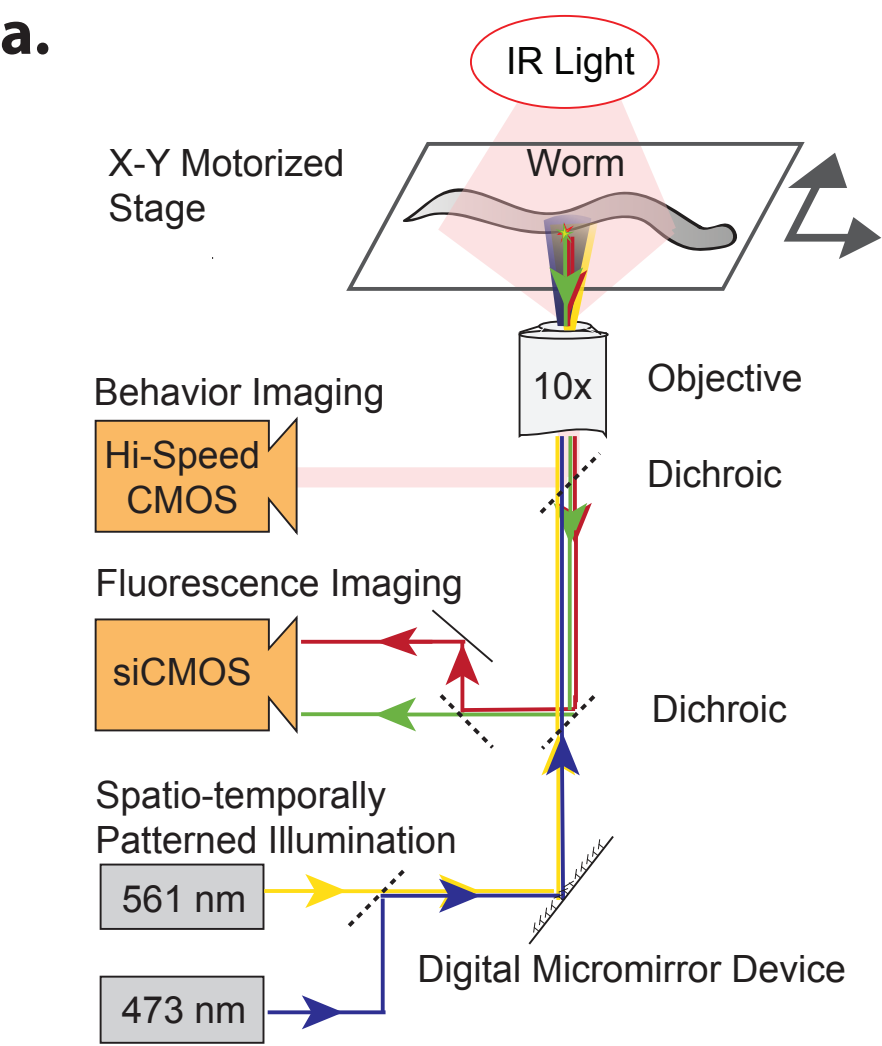
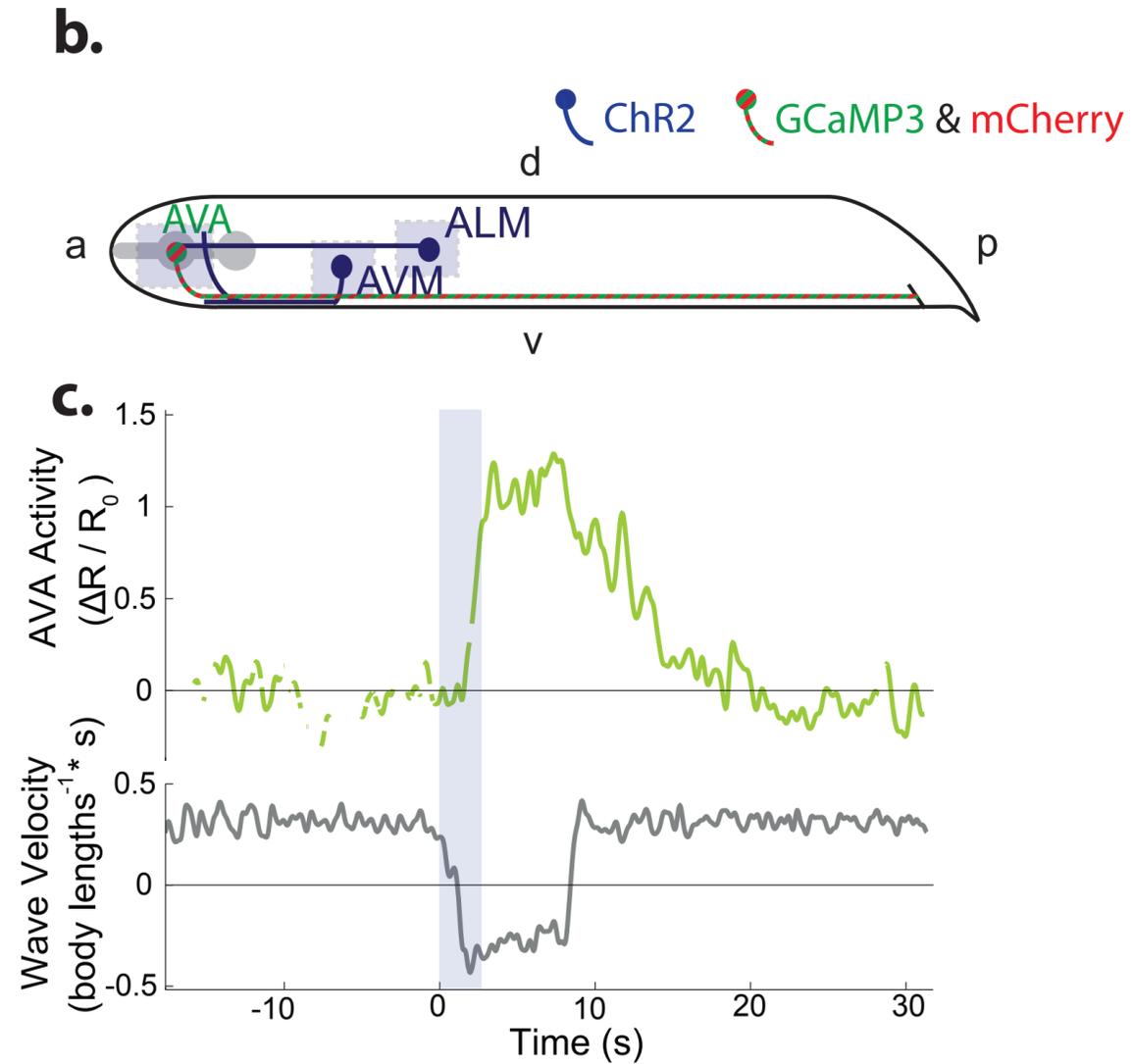
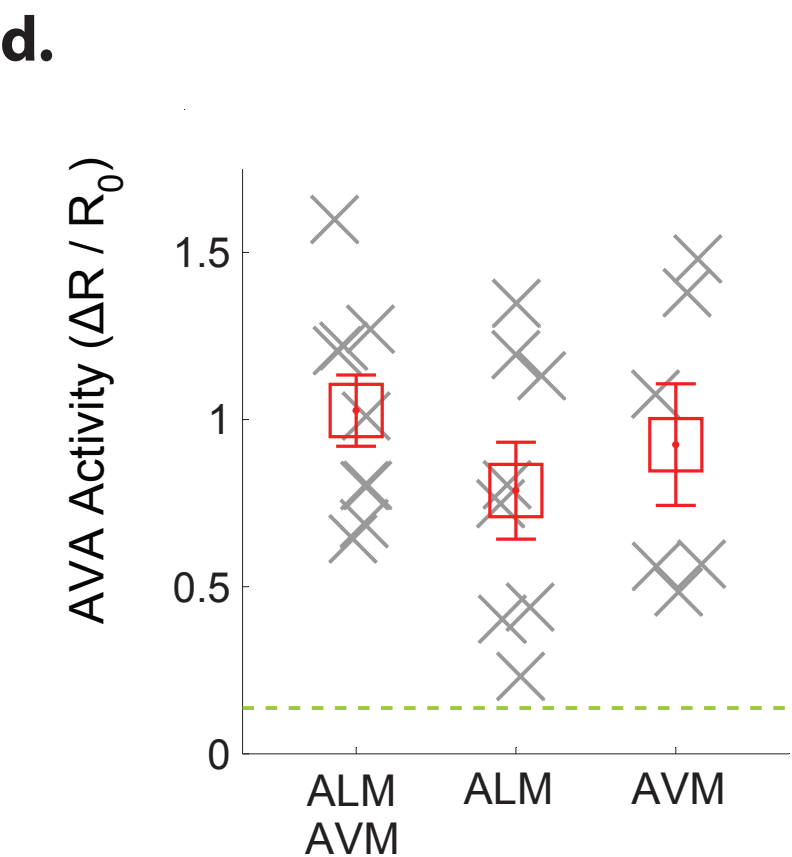
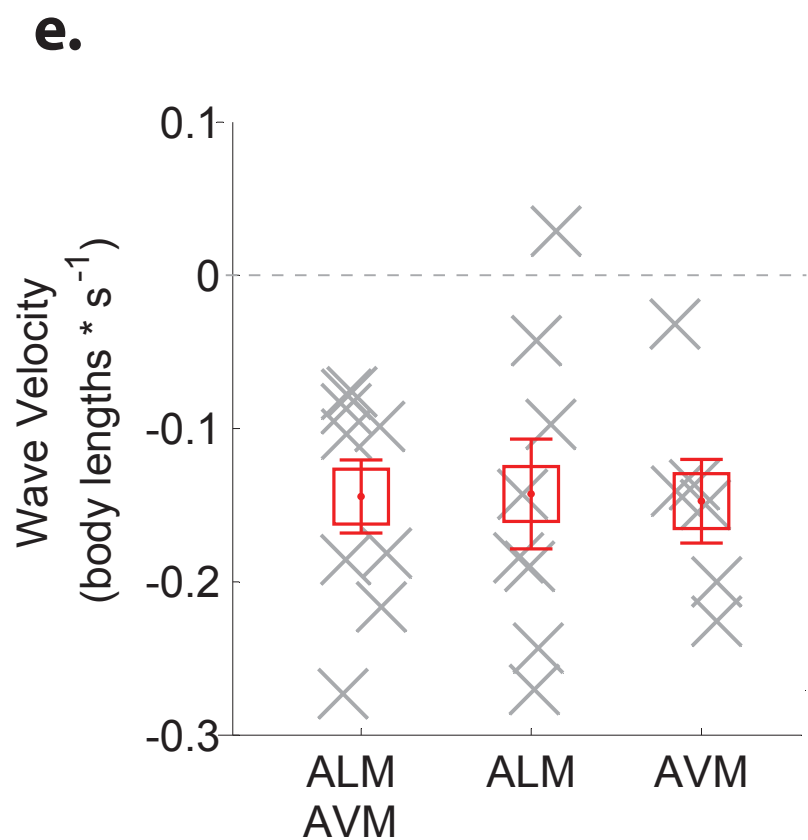
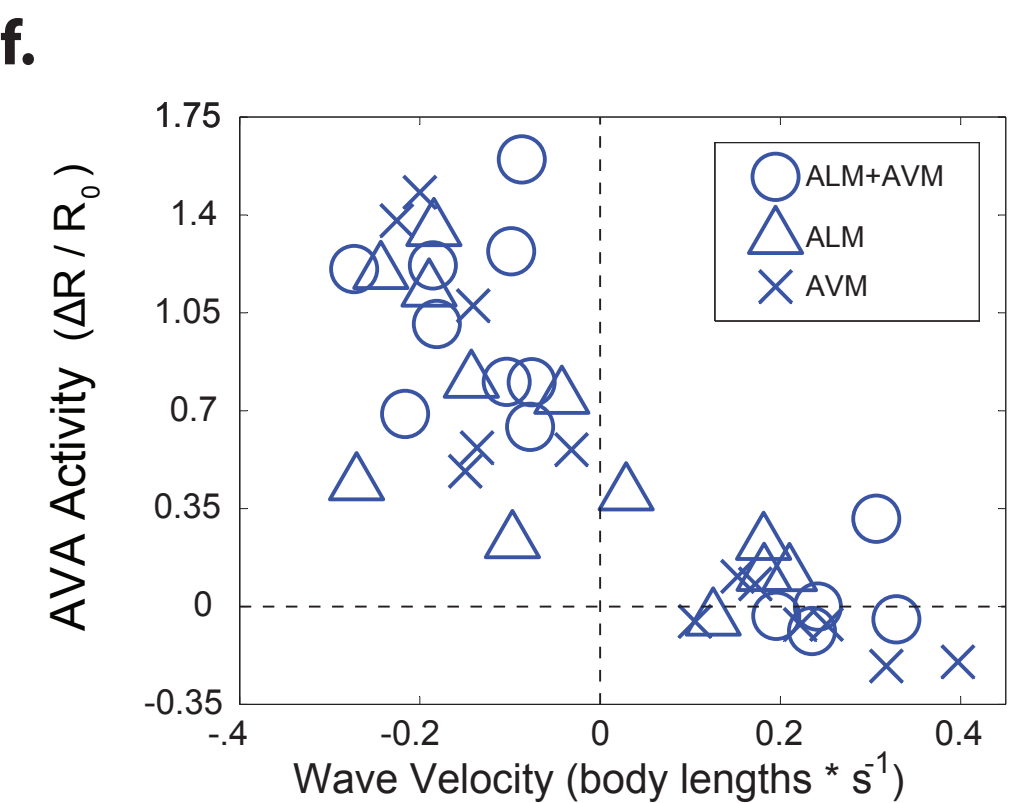
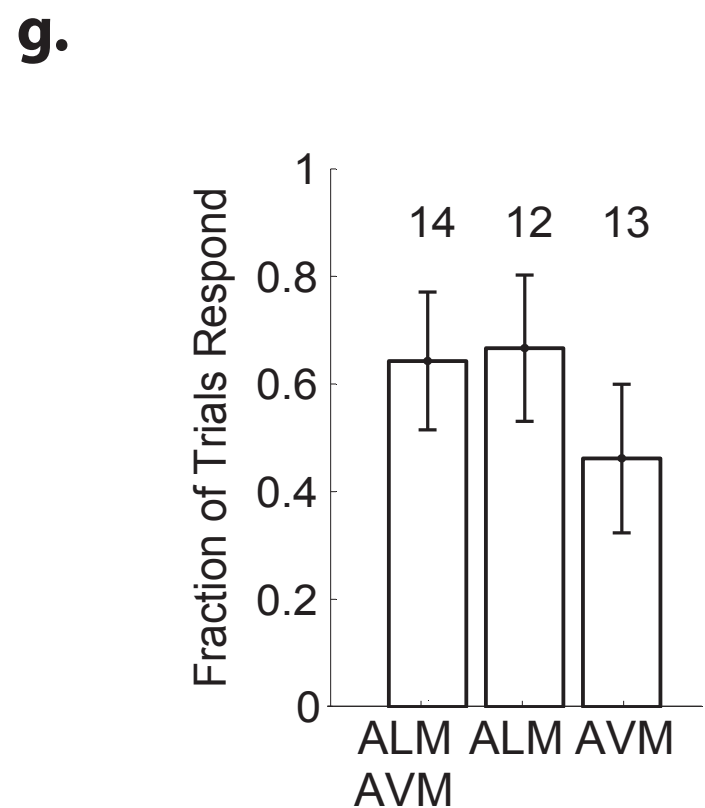

**Figure 1 Simultaneous optogenetic manipulation and calcium imaging in free-moving worms.**

**a.)** A digital micromirror device (DMD) reflects laser light onto only targeted neurons of a worm freely moving on a motorized stage. Real-time computer vision software monitors the worm's posture and location and controls the DMD, lasers and stage. **b.)** ChR2 is expressed in six soft-touch mechanosensory neurons, including ALM and AVM (dark blue). Calcium indicator GCaMP3 and fluorescent reference mCherry are expressed in command interneuron AVA (striped green and red). Light blue shaded regions indicate areas of illumination. "a" is anterior, "p" is posterior, "d" is dorsal and "v" is ventral. **c.)** Intracellular calcium dynamics of AVA (green line) are measured before during and after optogenetic stimulation of the ALM touch neuron (2.7 s stimulation, blue shaded region). The velocity of the worm's body bending waves is shown in gray. **d.)** Mean of activity and **e.)** velocity during a time window were shown (grey crosses) for worms that responded by reversing to optogenetic illumination of ALM ($n=8$ trials), AVM ($n=6$ trials) or both ($n=9$). Mean across trials is shown (red squares). Errorbars represent standard error of the mean. Dashed green line represents the mean plus standard deviation of fictive neural activity from a GFP control, which serves as a measure of noise in the instrument. **f.)** For each trial, mean AVA activity is plotted against mean velocity, ($n=39$ trials), including non-responders. **g)** The fraction, $f$, of responders are shown for each stimuli. Number, $n$, indicates total trials. Error bars are $\sigma_f = \sqrt{f(1-f)/n}$.

**Supplementary Methods**
The optogenetics portion of the instrument is based on the CoLBeRT system [1]. The calcium imaging portion builds upon concepts from an earlier imaging-only system, called DualMag [8, 9].

*Strains.*
Strain AML1 used in our experiment was made by crossing QW309 *lin-15(n765ts)*; (*zfIs18*) [P*mec-4::ChR2::YFP*] with QW280 (*zfis42*) P[*rig-3::GCaMP3::SL2::mCherry*]. Control experiments were performed with QW1075 (zfEx416) [P*rig3::GFP*:SL2::*mCherry*], *(zfis18)* [P*mec-4::GFP*].

*Microscopy.*
The optogenetic stimulation and calcium imaging system was built around a modified Nikon Eclipse Ti-U inverted microscope that contained two stacked filter cube turrets. All imaging was conducted using a 10x, numerical aperture (NA) 0.45, Plano Apo objective. Worms were imaged on 6cm NGM agarose plates on a Ludl BioPrecision2 XY motorized stage controlled by a MAC 6000 stage-controller. Custom real-time computer vision software kept the worm centered in the field of view via an automated feedback loop.

*Behavioral imaging.*
Worm behavior was imaged on a CMOS camera (Basler aca2000-300km) under dark-field illumination in the near-infrared (NIR). Dark-field illumination was chosen to provide high contrast between the worm and the agarose plate. NIR illumination was chosen to avoid cross talk with simultaneous fluorescence imaging and also to minimized inadvertent ChR2 activation. An NIR filter, Semrock #FF01-795/150, was mounted in the illumination path of a Nikon halogen lamp. Dark-field illumination was obtained by using a Ph3 phase ring to create an annular pattern of illumination. A custom filter cube (Semrock #FF670-SDi01) in the microscope's upper filter cube turret reflected the behavior imaging path out of the microscope through what would normally be its epifluorescence illumination pathway. From there, a telescope composed of two achromat doublet plano-convex lenses was used to form an image on the camera. A long-pass filter, Omega Optical #3RD710LP, prevented stray light from entering the camera. The Basler camera transferred images to a desktop computer via a BitFlow Karbon PCI Express x8 10-tap Full Camera Link framegrabber. Images were recorded via the MindControl software discussed below.

*Spatio-temporally patterned laser illumination.*
To stimulate ChR2 and to illuminate GCaMP3, we used a 473-nm diode-pumped solid state laser (DPSS) (CNI Laser MBL-III-473, 150-mW maximum power, OptoEngine). Similarly, to illuminate mCherry we used a 561-nm DPSS laser (Sapphire 561-150 CW CDRH, 150-mW maximum power, Coherent). Laser illumination entered the microscope from the bottom filter cube turret.

The beams from the 473-nm and 561-nm lasers were first aligned to a common beam path by a series of mirrors and a dichroic mirror (Semrock #FF518-Di01). The combined beam path was then expanded using a telescope of two plano-convex achromat doublet lenses and reflected by a 2-inch diameter mirror onto a 1,024 x 768 element digital micromirror device (Texas Instruments DLP, Discovery 4100 BD VIS 0.7-inch XGA, Digital Light Innovations). The patterned light from the digital micromirror device was imaged onto the sample via an achromat doublet that served as a tube lens, and a custom filter cube (Chroma #59022bs dichroic, Semrock #FF01-523/610 emission filter) in the microscope's lower filter cube turret. Light intensity measured at the sample was 2 mW * mm^-2 of 473-nm light and 1 mW * mm^-2 of 561-nm light.

*Fluorescence imaging.*
Red and green channel fluorescent images were recorded simultaneously with a Hamamatsu Orca Flash 4.0 siCMOS camera at 30 fps, with 33 ms exposure on a second Dell Precision T7600 desktop computer. To simultaneously image mCherry and GCaMP3 side-by-side we used a DV2 two-channel imager from Photometrics containing a custom filter set (Chroma #565dcxr dichroic, Semrock #FF01-609/54-25 red emission filter, Semrock #FF01-520/32-25 green emission filter). Fluorescent images were captured using HCImage software (Hamamatsu) running on a dedicated Dell Precision T7600 computer running Windows 7 with two Intel Xeon Quad Core 3.3 GHz processors and 49 GB of RAM.

*Real-Time computer vision software*
An improved version of the MindControl software [1], written in C, was used to generate patterned illumination and perform real-time feedback of the stage. The software was rewritten for 64-bit Windows 7. Additionally new features were added to give the user more options for performing timed stimulations and to give the user more control over stage feedback parameters. The overall stability of the software was also improved. The MindControl software was run on a Dell Precision T7600 computer running Windows 7 with two Intel Xeon Quad Core 3.3 GHz processors and 16 GB of RAM. Source code is released under the GNU General Public License and is available for download on GitHub at https://github.com/leiferlab/mindcontrol .

*Fluorescence imaging analysis*
Neural activity is reported as normalized deviations from baseline of the ratio between GCaMP3 and mCherry fluorescence, $\frac{\Delta R}{R_0} = \frac{R-R_0}{R_0}$, where $R = \frac{I_{GCaMP3} - background_{green}}{I_{mCherry} - background_{red}}$. We have chosen to report fluorescence intensity of GCaMP3 as a ratio to the calcium insensitive mCherry so as to better account for artifacts from the animal's motion. The baseline $R_0$ is defined as the mean of $R$ from trial onset to stimulus onset. The intensities $I_{GCaMP3}$ and $I_{mCherry}$ were measured as the median pixel intensity in the green and red channels, respectively, of the 40% brightest pixels of a circular region of interest (ROI) centered on the maximal intensity of a Gaussian smoothed image of the neuron AVA. The ROI was selected in each frame by custom MATLAB scripts and confirmed by the user. The local background, $background_{green}$ and $background_{red}$ in the green and red channels, respectively, were measured as the median pixel intensity in an annulus around the neuron. 2048 x 2048 pixel images were binned to 1024 x 1024 pixels. The ROI for each neuron was a circle of radius 8 pixels and the background was an annulus of inner radius 20 pixels and an outer radius of 22 pixels. At our magnification the scale was 0.62 μm / pixel. Custom MATLAB scripts were used to calculate the $\frac{\Delta R}{R_0}$ for each frame, and smoothed with a low-pass Gaussian filter (σ = 5 frames).

Transgenic animals expressing GFP instead of GCaMP3 were used as a control for motion artifacts and instrument noise. These animals were imaged under the same conditions as the GCaMP3 animals, however their reversals were spontaneous rather than induced.

*Optogenetic-induced behavior experiments*
Worms were grown on NGM agar plates seeded with 250μL of OP50 *E. Coli* mixed with 1μL of 1mM all-trans retinal in ethanol solution. Plates were seeded on day 0, worms were transferred to seeded plates on day 1 and imaging was performed on day 2.

Spatially distinct regions of the worm's body were defined to illuminate the cell bodies of selected neurons. The AVA region corresponded to 90% of the body width and 10% of the body length, centered 10% of the way from the anterior tip of the worm. The AVM region corresponded to 50% of the body width and 13% of the body length, centered 35.5% from the anterior tip, and 35% from the ventral edge. The ALM region was the same size, but centered 42.5% of the way from the anterior tip, and 35% from the lateral edge. Regions were selected to avoid overlap of processes, however the illumination region of AVA includes a small portion (estimated to be less then 20%) of the ALM process.

To mitigate suspected worm-to-worm variability in ChR2 expression and retinal uptake, we selected worms for our experiment that reversed in response to whole head illumination but did not reverse or pause in response to illuminating the small portion of the ALM process near AVA. Previously we had observed that illuminating small areas of neuronal processes alone usually caused little effect [1]. Approximately three quarters of worms tested responded with reversals to a brief whole head illumination, and of those, 39 of 53 worms had no noticeable response to the onset of process illumination and were thus deemed suitable for experimentation. Whole head illumination was performed manually on a fluorescent dissection scope (Nikon SMZ-1500) while process illumination was performed on the main instrument by illuminating the AVA region.

For imaging, worms were washed in M9, transferred to approximately 1.5-mm thick NGM agarose plates and covered with mineral oil to improve contrast under dark-field illumination. Imaging began after allowing animals to acclimate for 5 minutes. After >15 seconds of AVA imaging, either AVM, ALM or both, were illuminated for 2.7 seconds. AVA imaging continued for >15 seconds. Worms underwent up to two stimulations, with 3 minutes rest in between, before being discarded. Trials where the worm underwent spontaneous reversals prior to stimuli, or prolonged bouts of multiple distinct reversals in response to the stimuli were excluded.

*Data Analysis*
Behavioral data and calcium activity data were analyzed using custom MATLAB scripts, available at http:://github.com/leiferlab/dualmag-analysis. An empirical model of AVA activity was fit to each calcium trace to find the mean velocity and amplitude of neural activity for each trial. Calcium levels in AVA were assumed to increase and then decrease in a similar manner to an RC circuit in response to a voltage square wave,

$$f(t) = \begin{cases} 0 & 0 < t \\ A(1 - e^{-4t/\tau_1}) & \text{for } 0 < t < \tau_1, \\ Ae^{-(t-\tau_1)/(\tau_2-\tau_1)} & \tau_1 < t \end{cases}$$

where stimulus onset occurs at *t*=0, $A$ is the amplitude of neural activity and $\tau_1$ and $\tau_2$ are the timescales for calcium increase and decrease. Mean velocity was calculated over the time window from stimulus onset (t=0) to $\tau_1$. The parameters $A$, $\tau_1$ and $\tau_2$ were fit for each trace using least-mean squared and the Nelder-Mead method.

*Supplementary Methods References*
[7] Leifer, A.M. Optogenetics and computer vision for *C. elegans* neuroscience and other biophysical applications. Thesis. Harvard University, (2012).

[8] Clark, C.M., Leifer, A.M., Florman, J., Mizes, K.G., Samuel, A.D.T., Alkema, M.J. The coordianted activity of distinct sub-motorprograms give rise to a complex behavior, manuscript in preparation.

*Acknowledgements*
Thanks to Josh Shaevitz and Dmitri Aranov for productive discussions.

ALM_AVM

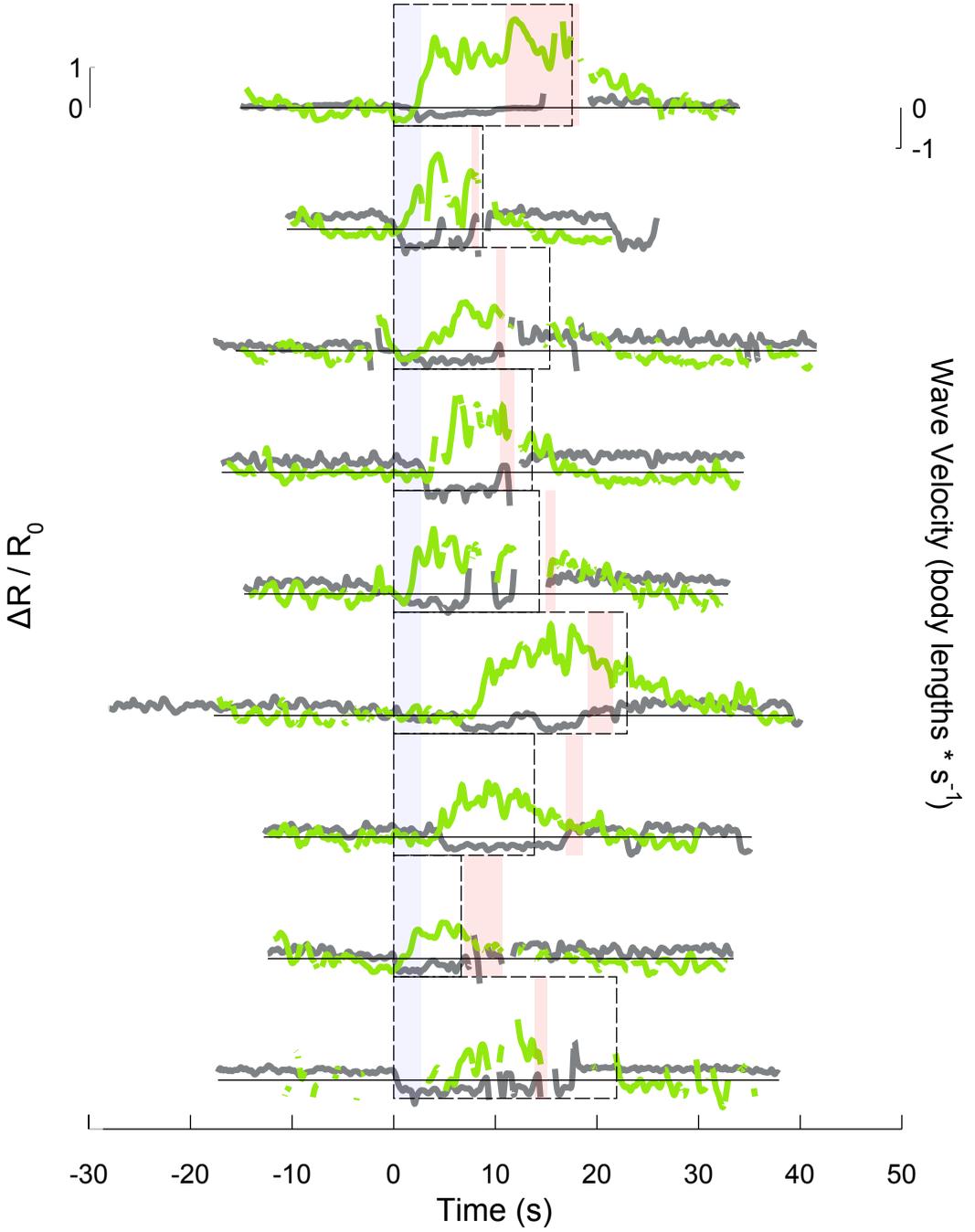

**Supplementary Figure 1**
Calcium dynamics of command interneuron AVA and animal velocity are shown before, during and after a 2.7 second stimulation to mechanosensory neurons ALM and AVM for each trial. All responding trials for this stimulation are shown. The transgenic animal shown here expresses GCaMP3 and mCherry in interneuron AVA and ChR2 in the soft-touch mechanosensory neurons (*lin-15(n765ts)*; (zfIs18) [P*mec-4*::*ChR2*::*YFP*]; (zfis42) P[*rig-3*::*GCaMP3*::*SL2*::*mCherry*]). Stimulus onset occurs at t=0 seconds (blue shaded region). Red shaded region indicates when the worm undergoes a deep ventral bend, called an "omega turn." Dashed rectangle indicates time region of interest over which velocity is averaged for each trial.

**Supplementary Video 1**
Calcium dynamics of command interneuron AVA are shown before, during and after a 2.7 second stimulation to mechanosensory neuron ALM. The transgenic animal shown here expresses GCaMP3 and mCherry in interneuron AVA and ChR2 in the soft-touch mechanosensory neurons (*lin-15(n765ts)*; (zfIs18) [P*mec-4*::*ChR2*::*YFP*]; (zfis42) P[*rig-3*::*GCaMP3*::*SL2*::*mCherry*]) . Upper left panel shows green fluorescence from GCaMP3 in false color. The upper middle panel shows red fluorescence from mCherry in false color. Orange circle indicates location of command interneuron AVA. The upper right panel shows worm behavior. Green triangle indicates head. Red square indicates tail. The animal's outline is shown in green. Blue area shows region of blue laser light illumination. Lower panel shows calcium activity. Stimulus onset occurs at t=0 seconds. The trial shown here is the same as that in Fig 1c.

**Supplementary Video 2**
A control worm expresses GFP in interneuron AVA and in the mechanosensory neurons including ALM and AVM ((*zfEx416*) [P*rig3*::*GFP*:Ssl2::*mCherry*], *(zfis18)* [P*mec-4*::*GFP*]). Different combinations of AVA, ALM and AVM are illuminated independently. Arrow indicates location of AVA. The text indicates targeted neuron. No fluorescence is observed in non-targeted regions.